\RequirePackage{lineno}
\documentclass[aps,prl,reprint,superscriptaddress,showpacs,showkeys,amsmath,amssymb,floatfix,nofootinbib]{revtex4-1}
\usepackage[colorlinks=true,citecolor=blue,filecolor=blue,linkcolor=blue,urlcolor=blue,pdftex]{hyperref}
\usepackage{booktabs,graphicx,mathrsfs,verbatim,amsmath,units,soul}
\usepackage[colorlinks=true,citecolor=blue,filecolor=blue,linkcolor=blue,urlcolor=blue,pdftex]{hyperref}
\usepackage[usenames,dvipsnames]{color}
\usepackage{ulem}
\usepackage{color}
\usepackage{upgreek}

\newcommand\figref[1]{Fig.~\ref{#1}}

\begin{document}	
\title{Search for Electronic Recoil Event Rate Modulation with 4 Years of XENON100 Data}

\newcommand{\freiburg}{\affiliation{Physikalisches Institut, Universit\"at Freiburg, 79104 Freiburg, Germany}}
\newcommand{\bologna}{\affiliation{Department of Physics and Astrophysics, University of Bologna and INFN-Bologna, 40126 Bologna, Italy}}
\newcommand{\chicago}{\affiliation{Department of Physics \& Kavli
Institute of Cosmological Physics, University of Chicago, Chicago, IL 60637, USA}}
\newcommand{\coimbra}{\affiliation{Department of Physics, University of Coimbra, 3004-516, Coimbra, Portugal}}
\newcommand{\columbia}{\affiliation{Physics Department, Columbia University, New York, NY 10027, USA}}
\newcommand{\lngs}{\affiliation{INFN-Laboratori Nazionali del Gran Sasso and Gran Sasso Science Institute, 67100 L'Aquila, Italy}}
\newcommand{\mainz}{\affiliation{Institut f\"ur Physik \& Exzellenzcluster PRISMA, Johannes Gutenberg-Universit\"at Mainz, 55099 Mainz, Germany}}
\newcommand{\heidelberg}{\affiliation{Max-Planck-Institut f\"ur Kernphysik, 69117 Heidelberg, Germany}}
\newcommand{\munster}{\affiliation{Institut f\"ur Kernphysik, Wilhelms-Universit\"at M\"unster, 48149 M\"unster, Germany}}
\newcommand{\nikhef}{\affiliation{Nikhef and the University of Amsterdam, Science Park, 1098XG Amsterdam, Netherlands}}
\newcommand{\nyuad}{\affiliation{New York University Abu Dhabi, Abu Dhabi, United Arab Emirates}}
\newcommand{\purdue}{\affiliation{Department of Physics and Astronomy, Purdue University, West Lafayette, IN 47907, USA}}
\newcommand{\rpi}{\affiliation{Department of Physics, Applied Physics and Astronomy, Rensselaer Polytechnic Institute, Troy, NY 12180, USA}}
\newcommand{\rice}{\affiliation{Department of Physics and Astronomy, Rice University, Houston, TX 77005, USA}}
\newcommand{\stockholm}{\affiliation{Oskar Klein Centre, Department of Physics, Stockholm University, AlbaNova, Stockholm SE-10691, Sweden}}
\newcommand{\subatech}{\affiliation{SUBATECH, Ecole des Mines de Nantes, CNRS/In2p3, Universit\'e de Nantes, Nantes 44307, France}}
\newcommand{\torino}{\affiliation{INFN-Torino and Osservatorio Astrofisico di Torino, 10125 Torino, Italy}}
\newcommand{\ucla}{\affiliation{Physics \& Astronomy Department, University of California, Los Angeles, CA 90095, USA}}
\newcommand{\ucsd}{\affiliation{Department of Physics, University of California, San Diego, CA 92093, USA}}
\newcommand{\wis}{\affiliation{Department of Particle Physics and Astrophysics, Weizmann Institute of Science, Rehovot, 7610001, Israel}}
\newcommand{\zurich}{\affiliation{Physik-Institut, University of Zurich, 8057 Zurich, Switzerland}}

\author{E.~Aprile}\columbia
\author{J.~Aalbers}\nikhef
\author{F.~Agostini}\lngs\bologna
\author{M.~Alfonsi}\mainz
\author{F.~D.~Amaro}\coimbra
\author{M.~Anthony}\columbia
\author{F.~Arneodo}\nyuad
\author{P.~Barrow}\zurich
\author{L.~Baudis}\zurich
\author{B.~Bauermeister}\stockholm
% ...
\collaboration{XENON Collaboration}\email[E-mail: ]{xenon@lngs.infn.it}\noaffiliation

\date{\today}

\begin{abstract} 
We report on a search for electronic recoil event rate modulation signatures in the XENON100 data accumulated over a period of 4 years, from January 2010 to January 2014. A profile likelihood method, which incorporates the stability of the XENON100 detector and the known electronic recoil background model, is used to quantify the significance of periodicity in the time distribution of events. There is a weak modulation signature at a period of $431^{+16}_{-14}$~days in the low energy region of $(2.0-5.8)$~keV in the single scatter event sample, with a global significance of $1.9\,\sigma$, however no other more significant modulation is observed. The expected annual modulation of a dark matter signal is not compatible with this result. Single scatter events in the low energy region are thus used to exclude the DAMA/LIBRA annual modulation as being due to dark matter electron interactions via axial vector coupling at $5.7\,\sigma$.

\end{abstract}
\maketitle

The DAMA/LIBRA experiment has reported the observation of a periodic annual modulation of the low-energy~(low-E), ($2-6$)~keV, single-hit event rate in their NaI 
detectors~\cite{Bernabei:dama}. The interpretation of this modulation as being 
due to WIMP-induced nuclear recoils (NRs) has been challenged by null 
results from several other experiments using different target materials and detector 
technologies~\citep[e.g.][]{Tan:pandax,Aprile:2016run12,Agnese:cdmslowe,Akerib:lux,Amole:pico,Agnes:darkside}, 
most of which have considerably lower radioactive backgrounds.
There are several alternative theories predicting dark matter (DM)-induced electronic recoils (ERs) as an explanation 
for the DAMA/LIBRA modulation~\citep[e.g.][]{Kopp:leptonic}. These hypotheses, 
however, have also been challenged by results from XENON100 using ER data~\cite{Aprile:dcpaper,Aprile:acpaper} and XMASS using ER/NR-agnostic data similarly to DAMA/LIBRA~\cite{Abe:xmass}.

The XENON100 detector is a dual phase xenon time projection chamber (TPC) that measures the direct scintillation (S1) and delayed proportional scintillation light (S2) from a particle interacting in the liquid xenon (LXe)~\cite{Aprile:2012instr}. Information such as the energy and position of the interaction can be reconstructed from the S1 and S2 signals. In this analysis, we combine the three science runs of XENON100 to further test the 
hypothesis of DM inducing ER. Run~I lasted from January~13,~2010 to 
June~8,~2010~\cite{Aprile:2012run08}, and Run~II from February~28,~2011 to March~31,~2012. Run~II was previously used to search for 
spin-independent (SI)~\cite{Aprile:2012run10} and spin-dependent (SD)~\cite{Aprile:2013sd} 
WIMP-induced NRs, axion-induced ERs~\cite{Aprile:axion}, 
and to test DAMA/LIBRA using the average~\cite{Aprile:dcpaper} 
and time-dependent~\cite{Aprile:acpaper} ER rates. Run~III data was accumulated from 
April~22,~2013 to January~8,~2014 and was combined with the previous two runs  
to update the SI and SD analyses~\cite{Aprile:2016run12}. 
The three runs have 100.9, 223.1, and 153.0 live-days, respectively, and together span a total of 1456 calendar days ($\simeq4$~years).

The ER energy reconstructed from S1 and the uncertainty herein are determined as in~\cite{Aprile:axion,Aprile:acpaper}. The low-E range ($3-14$)~PE corresponds 
to ($2.0-5.8$)~keV and thus covers the energy interval where DAMA/LIBRA 
observes an annual modulation. The high energy (high-E) range 
($14-30$)~PE corresponds to ($5.8-10.4$)~keV and is used as a side band control sample. 

\begin{figure*}
\centering
\includegraphics[width=1\textwidth]{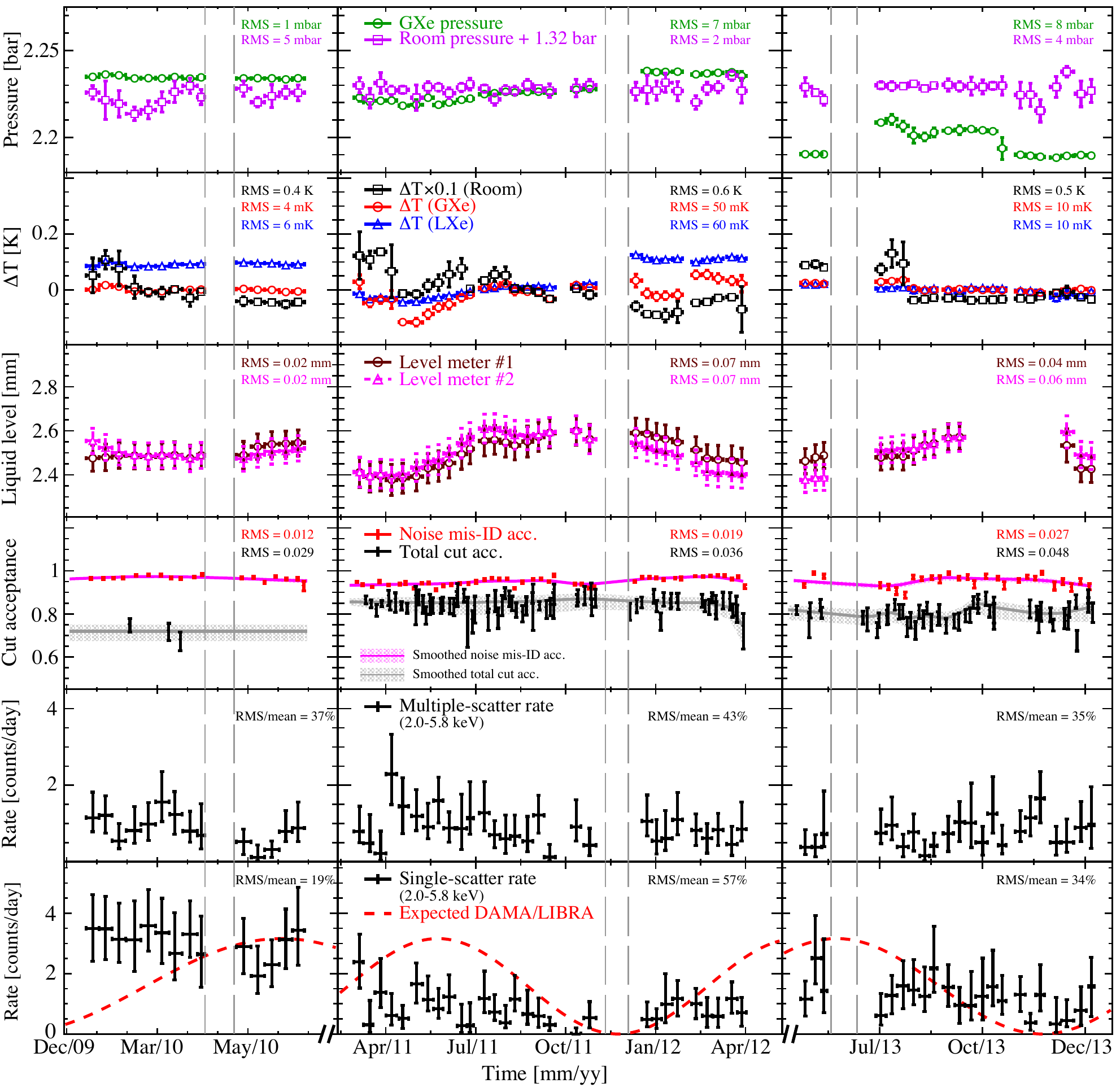}
\vspace{-0.2in}
\caption{\label{stability} 
Temporal evolution of the most relevant quantities across Run~I (left column), Run~II (middle column), and Run~III (right column), spanning a range of 4~years. The vertical dashed lines indicate detector maintenance periods. The panels (from top to bottom) show the detector/room pressures, various temperature readings, height of the liquid xenon level, signal acceptances, and low-E MS and SS event rates (acceptance corrected). The uncertainty bands on the acceptance models are derived from the weighted mean of the $1\sigma$ error bars. The expected DAMA/LIBRA modulation signal (assuming a 100\% modulation fraction) in the XENON100 detector as calculated in the text is overlaid in the bottom panel for comparison.
}
\vspace{-0.1in}

\end{figure*}

Low-E single scatter (SS) events in the 34~kg fiducial mass, as expected from DM interactions, are selected using the same criteria as in the respective DM search analysis for that run (Run~I~\cite{Aprile:2012run08}, Run~II~\cite{Aprile:2012run10, Aprile:xe100analysis}, Run~III~\cite{Aprile:2016run12}). While these criteria are defined to select valid NR events, they also have high efficiency for ERs~\cite{Aprile:axion,Aprile:dcpaper, Aprile:acpaper}. Low-E multiple scatter (MS)
events, selected as SS events in the fiducial volume with a coincident S1 in the active veto surrounding the LXe TPC, are used as a second control sample. The acceptances in each energy range are derived following the procedure in~\cite{Aprile:xe100analysis} using weekly ER calibration data 
($^{60}$Co and $^{232}$Th). This takes into account the acceptance loss due to the misidentification of correlated electronic noise as an S1 signal (``noise mis-ID''), described in~\cite{Aprile:2016run12}. A new data quality cut removes exceptionally noisy datasets where the acceptance loss due to noise mis-ID is $>0.1$ in the low-E region of (3-14)~PE, resulting in a total livetime reduction of 18.0 (26.7) days in Run~II~(III). The time variation of the acceptances in the low-E range, shown in \figref{stability}, are incorporated in the analysis by smoothly interpolating between the data points. Adopting different methods of interpolation do not significantly affect the results. 

The stability of the XENON100 detector is studied through various characteristic 
parameters such as liquid xenon level, pressures, and temperatures of gaseous and liquid xenon monitored by sensors distributed within the system. The parameters with the highest potential impact on detector 
signals are shown in Fig~\ref{stability}. The relative fluctuations in the pressures, temperatures, and liquid level are less than 2\% during each run. 

Small variations of the detector parameters may influence signal generation inside the detector, potentially affecting acceptances and event rates.
Thus, linear (Pearson) and non-linear (Spearman-Rank) correlation coefficients between the
detector parameters and SS and MS event rates in each energy range are calculated to identify potential correlations. Different run conditions 
cause the offsets in the parameters between runs in \figref{stability}. To avoid 
artificial correlations from these offsets, the detector parameters are normalized 
across all runs prior to the calculation of the correlation coefficients.
Uncertainties in both the detector parameters and event rates are taken into account 
through multiple pseudo-experiments, in which the data points are sampled based on the 
error bars in \figref{stability}.
No significant correlations with p-values smaller than 0.1 are found between event rate and any detector parameter, 
which suggests that the correlation with detector temperature~($>2.8\sigma$) and liquid level~($>2.5\sigma$) observed 
in the previous Run~II-only analysis~\cite{Aprile:acpaper} was coincidental. Several binning configurations have been tested, resulting in the same conclusion of no correlation between detector and background rates.

Variations in the background from external $\gamma$ and internal $\beta$ radiation 
can affect the search for event rate modulations. Of all external $\gamma$ sources, 
only $^{60}$Co ($\tau_{\rm{Co}}=7.6~y$) decays fast enough to cause an observable change in event rate over the 4 years time range considered here. The contributions to the SS and MS event rates on January~1, 2011, $P_{\rm{Co}}$, are $(0.47\pm0.02)$ and $(1.76\pm0.03)$~events/(keV$\cdot$tonne$\cdot$day), respectively, estimated by Monte Carlo simulation using measured material contamination as input~\cite{Aprile:ermc}. 

\begin{figure*}[!htp]
\centering
\includegraphics[width=1\textwidth]{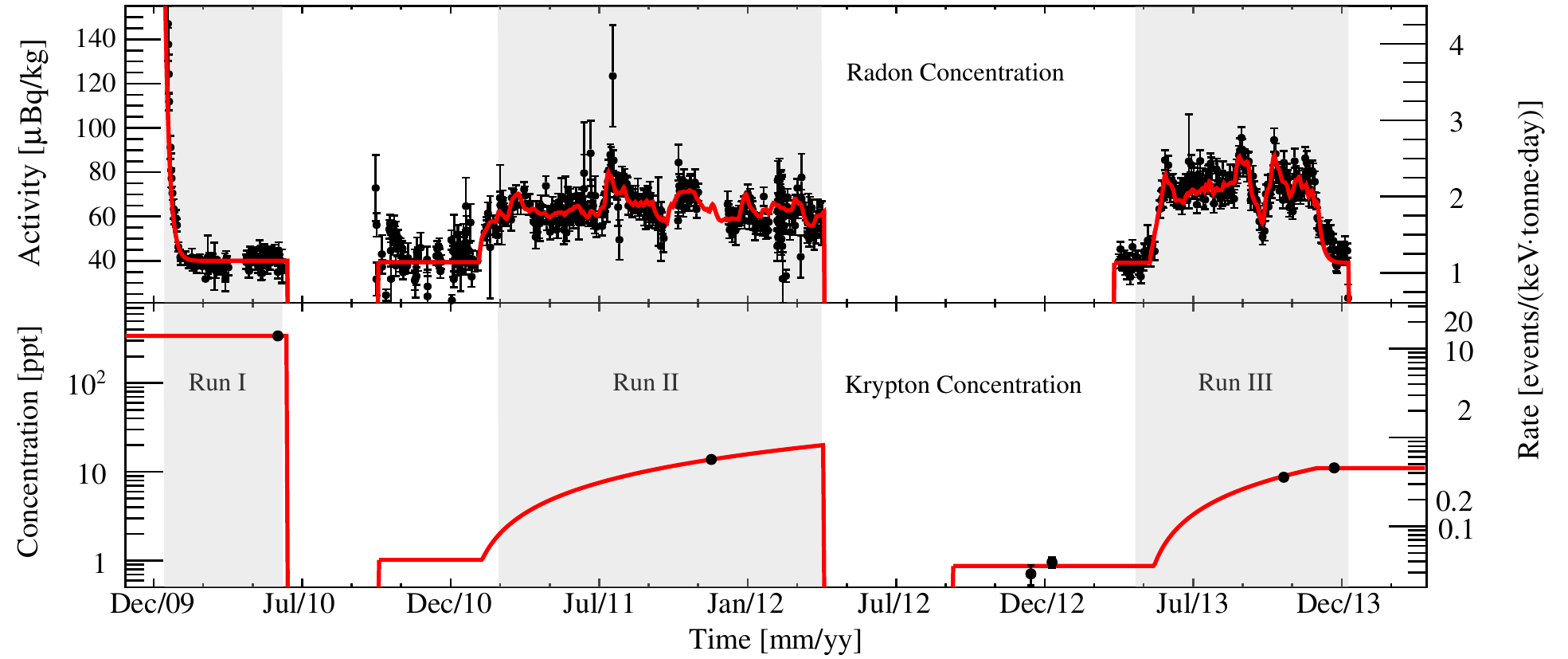}
\vspace{-0.2in}
\caption{\label{radon_krypton}
Temporal evolution of (top) $^{222}$Rn activity via characteristic alpha spectroscopy and (bottom) krypton concentration via RGMS measurement in the XENON100 detector. The $^{222}$Rn evolution is modeled by Eq.~(\ref{eq:radon_correlation}) and shown by the red line in the top panel. The $^{85}$Kr concentration is modeled by linearly increasing functions during air leaks in Run~II and Run~III, and constant otherwise due to its long decay lifetime of $\tau_{\rm{Kr}}=15.6$~years, as shown by the red line in the bottom panel. The shaded regions show the range of each run.
}
\vspace{-0.1in}
\end{figure*}

Under nominal conditions, the radon level inside the LXe is given by the emanation of detector and circulation loop surfaces and should thus be constant in time. The same holds for the background from decays of $^{85}$Kr. However, tiny air leaks at the diaphragm pump used for xenon circulation were identified, which led to a time-variable radon and krypton background. The $^{222}$Rn activity in the detector was monitored via characteristic alpha and $\beta-\gamma$ delayed coincidences (\figref{radon_krypton} top). A correlation analysis with the measurement of the radon activity in the laboratory ($C^{ext}_{\rm{Rn}}$) provides a model of the time evolution of the $^{222}$Rn concentration as

\begin{equation}
\centering
C_{\rm{Rn}}^{in}(t) = C^{in}_{\rm{const}} + L  \int_{-\infty}^{t} C_{\rm{Rn}}^{ext}(t') 
\times e^{\left(\frac{t'-t}{\tau_{\rm{Rn}}}\right)} \rm{d}t',
\label{eq:radon_correlation}
\end{equation}
where $C^{in}_{\rm{const}}$ is the $^{222}$Rn activity from emanation and $L$ is the air leak rate. The model describes the measured data very well (\figref{radon_krypton} top), with good agreement between the constant activities in each of the three runs as expected. The activity is translated to a low-E SS event rate by scaling of $P_{\rm{Rn}}=(0.029\pm0.002)$~(events/(keV$\cdot$tonne$\cdot$day))/$(\upmu$Bq/kg)~\cite{Aprile:ermc}, where the uncertainty is dominated by the measured reduction of the $^{218}$Po level compared to the original $^{222}$Rn.

The time evolution of air leaks described by this model is also used to model the time dependence of the $^{\rm{85}}$Kr background, for which fewer direct measurements exist via offline rare gas mass spectroscopy (RGMS)~\cite{rgms,sebastianthesis}. The resulting model for the $^{\rm{nat}}$Kr concentration agrees very well with the measurements (\figref{radon_krypton} bottom). The contribution of $^{85}$Kr to low-E SS events is determined as $R_{\rm{Kr}}=P_{\rm{Kr}}\cdot C_{\rm{Kr}}$, with a conversion coefficient $P_{\rm{Kr}} =(4.1\pm0.8)\times10^{-2}$~(events/(keV$\cdot$tonne$\cdot$day))/ppt~\cite{Aprile:ermc}. The uncertainty is from the measured $^{85}$Kr to $^{\rm{nat}}$Kr ratio of $(2.1\pm0.3)\times10^{-11}$~\cite{sebastianthesis} and systematic uncertainties from the RGMS measurements~\cite{rgms}. Background contributions from $^{60}$Co, $^{222}$Rn and $^{85}$Kr are all taken into account in the statistical analysis presented below.

The statistical significance of a potential modulation signal is determined by an unbinned profile 
likelihood (PL) method as in~\cite{Aprile:acpaper}. In the presence of a modulation signal, the event rate for each run $i$ is modeled as
\begin{widetext} 
\begin{equation}
\centering
\begin{aligned}[width=1\textwidth]
f_i(t)=\epsilon^1_i(E,t,P_{\epsilon, i}^{1})\cdot\epsilon^2_i(E,t,P_{\epsilon, i}^{2})\cdot\left[C+A\cdot\cos\left(2\pi \cdot \frac{(t-\phi)}{P}\right)+P_{\rm{Co}}\cdot e^{-t/\tau_{\rm{Co}}}+P_{\rm{Rn}}\cdot C^{in}_{\rm{Rn}}(t)+P_{\rm{Kr}}\cdot C_{\rm{Kr}}(t)\right],
\end{aligned}
\label{eq:er_rate_pdf}
\end{equation}
\end{widetext}
where $\epsilon^{1,2}_i$ are the smoothed signal acceptances shown in \figref{stability},
$P_{\epsilon, i}^{1,2}$ are
nuisance parameters that scale each acceptance according to the uncertainty bands in \figref{stability},
$C$ is a constant event rate which includes both potential signal and the stable ER background, and a modulation signal is characterized by an amplitude $A$, with a period $P$, and phase $\phi$. 
The background-only hypothesis is described by Eq.~(\ref{eq:er_rate_pdf}) with $A = 0$.
Eq.~(\ref{eq:er_rate_pdf}) is normalized for each run to take into account the time distribution of data to become the probability density function, $\tilde{f}_i(t)$. The likelihood function is then constructed as
\begin{widetext}
\begin{equation}
\centering
\begin{aligned}
\mathcal{L} & =\prod^{\rm{Run~I,II,III}}_{i} \left[Poiss\left(n_{i}|N^{i}_{exp}(\eta)\right)\times\prod_{l=1}^{n_{i}}{\tilde{f}_i(t_{l},P_{\epsilon, i}^1,P_{\epsilon, i}^2,P_{\rm{Co}},P_{\rm{Rn}},P_{\rm{Kr}},A,\phi,P)}    \times {\mathcal{G}(P_{\epsilon, i}^{1})}   \times {\mathcal{G}(P_{\epsilon, i}^{2})}  \right]\\
  &  \times \mathcal{G}(P_{\rm{Co}})   \times \mathcal{G}(P_{\rm{Rn}})   \times \mathcal{G}(P_{\rm{Kr}})  \times \mathcal{G}(\eta),  
\end{aligned}
\label{eq:er_rate_likelihood}
\end{equation}
\end{widetext}
where $n_{i}$ and $N^{i}_{exp}$ are the total number of observed and expected events, respectively. Nuisance parameters are constrained by Gaussian penalty 
terms~$\mathcal{G}$, with the corresponding uncertainties discussed above. The parameters of interest are $P$, $A$ and $\phi$, while the other nuisance parameters are profiled out in the PL analysis.

\begin{figure}[h]
\centering
\includegraphics[width=1\columnwidth]{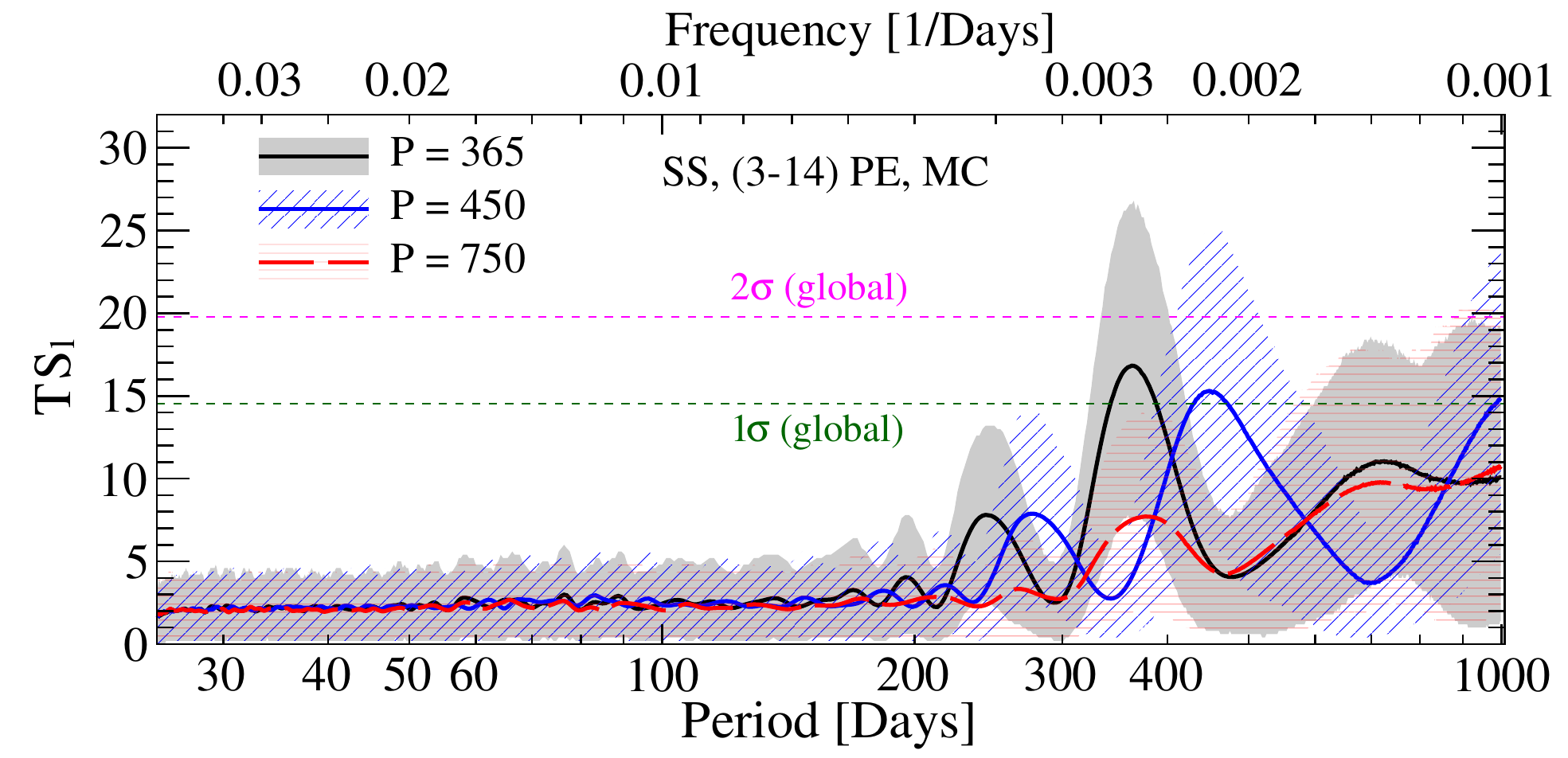}
\vspace{-0.2in}
\caption{\label{plres} The expected mean (solid lines) and central 68.3\% region (shaded bands) of test statistics as a function of period for simulated data. Uncertainties on all parameters are taken into account. The horizontal global significance lines are derived from the null hypothesis tests and shown here for comparison to \figref{plspecdata}.\vspace{-0.1in}
}
\end{figure}

The maximum profiled likelihoods are denoted by 
$\mathcal{L}_{0}$ for the null hypothesis and $\mathcal{L}_{1}$ for the modulation hypothesis. The local test statistics ($\rm{TS}_{\rm{l}}$) defined as $-2\,\rm{ln} (\mathcal{L}_{0}/\mathcal{L}_{1})$ and global test statistics ($\rm{TS}_{\rm{g}}$) are constructed in the same way as in \cite{Aprile:acpaper} to quantify the significance of a modulation signature. A Monte Carlo (MC) simulation based on Eq.~(\ref{eq:er_rate_pdf}), including all nuisance parameter variations, is used to evaluate the asymptotic distributions of the test statistics and to assess the sensitivity of the combined data to event rate modulations. The average test statistics of three representative periods with the same amplitude as in~\cite{Aprile:acpaper} are shown in \figref{plres} for SS samples in the low-E range. As the signal period increases, the resolution on the reconstructed period decreases and approaches a characteristic plateau above $\sim750$~days. As a result, the region of interest is restricted from 25 to 750 days in the PL analysis. In the Run~II-only analysis~\cite{Aprile:acpaper}, this plateau became apparent at $\sim500$~days. The side lobes next to the peak at each period are due to the time gaps between each run, verified by dedicated simulations.
 
\begin{figure}
\centering
\includegraphics[trim={0 1.43cm 0 0},clip,width=1\columnwidth]{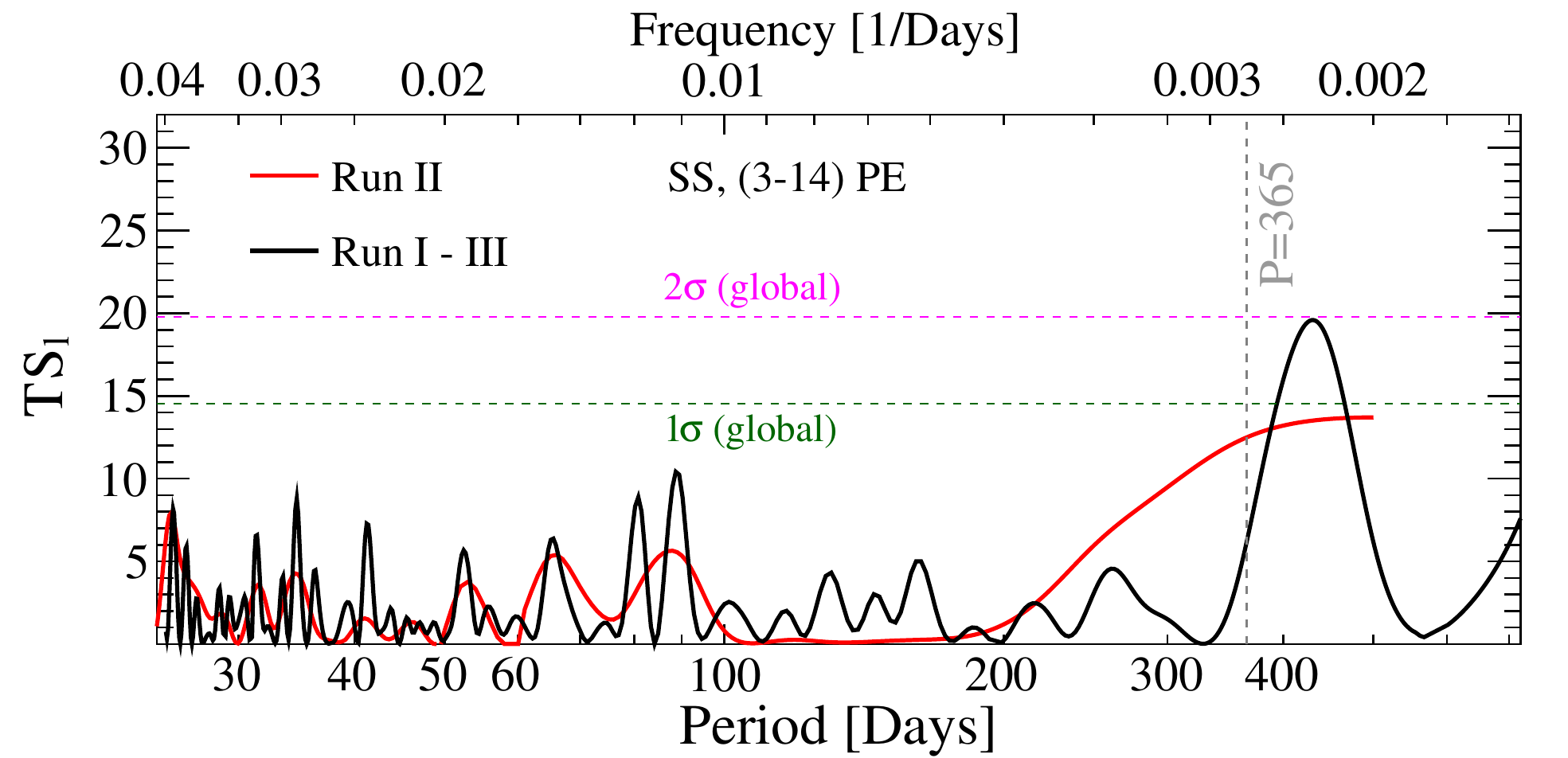}
\includegraphics[trim={0 1.43cm 0 1.44cm},clip,width=1\columnwidth]{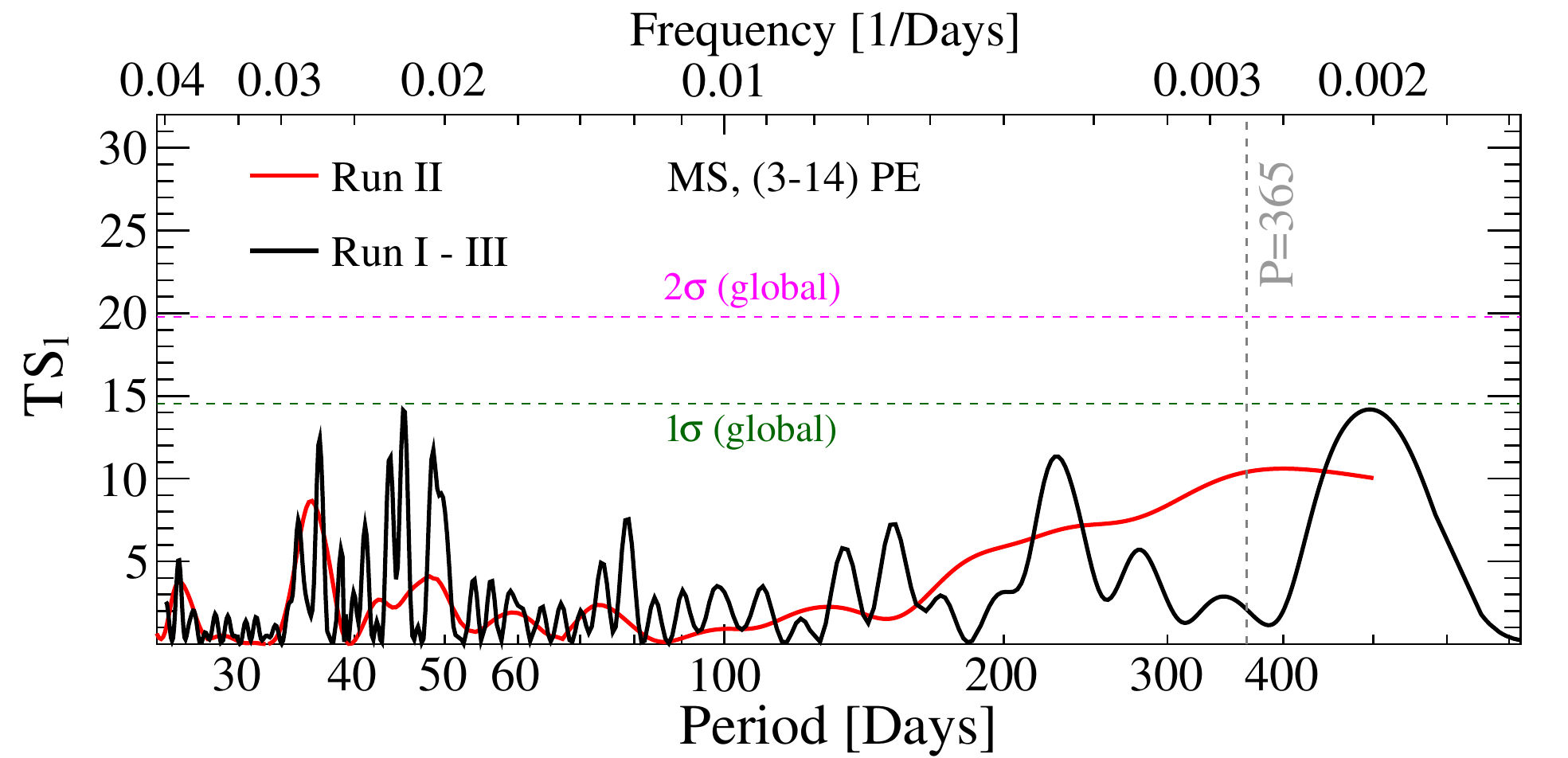}
\includegraphics[trim={0 0 0 1.44cm},clip,width=1\columnwidth]{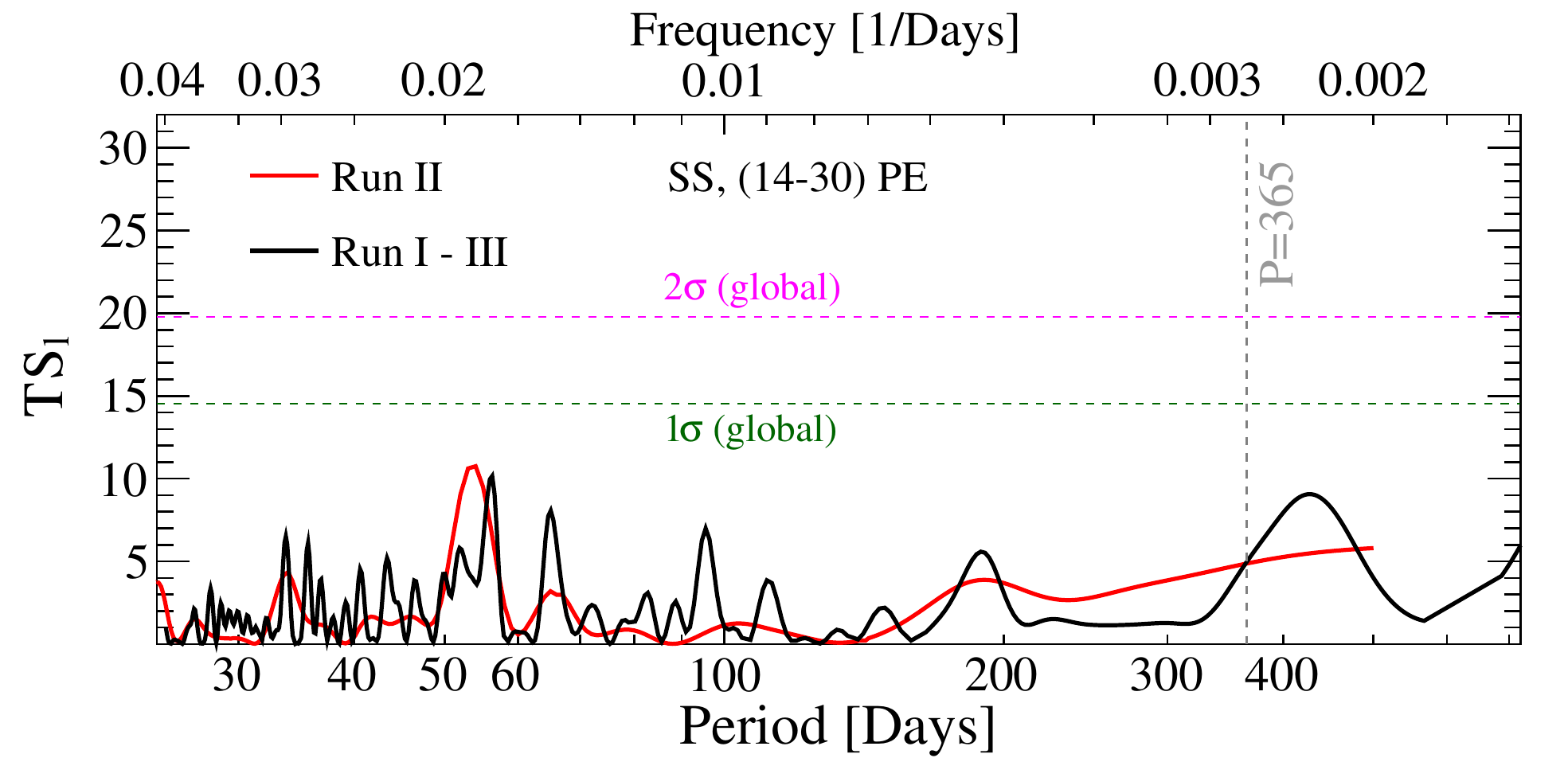}
\vspace{-0.2in}
\caption{\label{plspecdata} Test statistics as a function of modulation period for single scatters in the low-E region (top), multiple scatters in the low-E region (middle) and single scatters in the high-E region (bottom). The phase is unconstrained. The previous Run~II-only result~\cite{Aprile:acpaper} is overlaid for comparison.
\vspace{-0.1in}
}
\end{figure}

The PL results for the low-E SS signal sample and the two control samples are shown in \figref{plspecdata}. As the sensitivity and resolution increase by adding Run~I and Run~III data, the rising significance for the signal sample at large periods evident in Run~II data~\cite{Aprile:acpaper} becomes a distinguishable peak at $P=431^{+14}_{-16}$~days, reaching a global significance of $1.9\,\sigma$. The local significance for an annual modulation drops from $2.8\,\sigma$~\cite{Aprile:acpaper} to $1.8\,\sigma$, or to $2.2\,\sigma$ when fixing $\phi = 152$~days from the standard halo model. 

A similar peak at $495^{+32}_{-29}$~days period is indicated by the MS control sample, but the global significance is only $0.9\,\sigma$. The significance for annual modulation decreases to $0.4\,\sigma$. The shape of the significance spectrum in the high-E control sample is similar to the signal sample, but the peaks are not as evident. The similarity of the spectra between the two control samples and the signal sample further disfavors the possibility that the weak modulation signature indicated by this data is caused by DM interactions. 

\begin{figure}[th]
\centering
\includegraphics[width=1\columnwidth]
{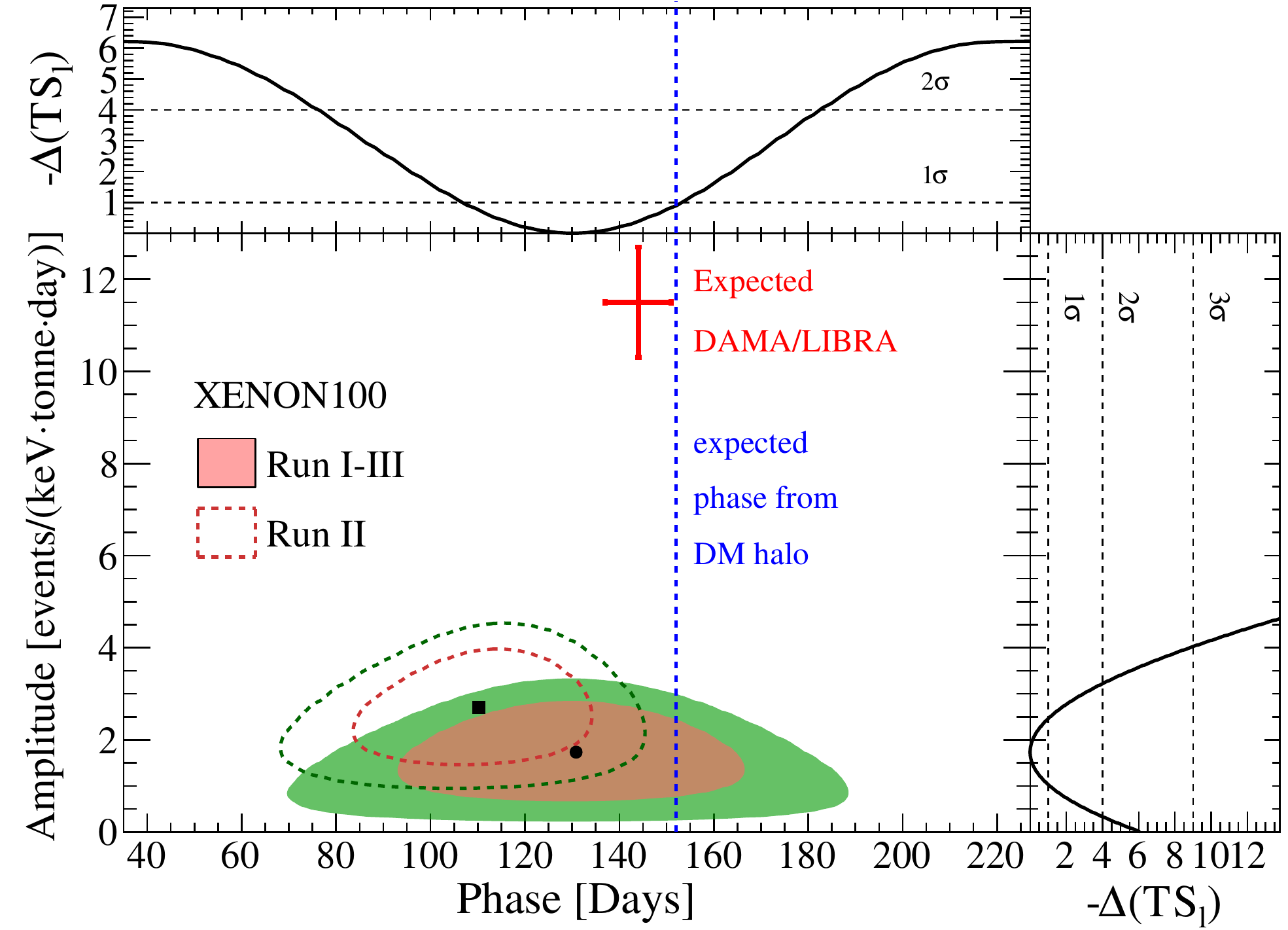}
\vspace{-0.2in}
\caption{\label{plphase} The XENON100 best-fit black dot, and $68\%$ (light red shaded region) and $90\%$ (green shaded region) confidence level contours as a function of amplitude and phase relative to January 1, 2011 for one year period. The corresponding Run~II-only results~\cite{Aprile:acpaper} are overlaid with a black square and dotted lines. The phase is less constrained than in Run~II due to the smaller amplitude. The expected DAMA/LIBRA signal (cross, statistical uncertainty only) and the phase expected from a standard DM halo (vertical dotted line) are shown for comparison. Top and side panels show $-\Delta(\rm{TS_{\rm{l}}})$ as a function of phase and amplitude, respectively, along with two-sided significance levels. 
}
\end{figure}

In absence of a significant annual modulation signature in the signal sample, the data is used to constrain the interpretation of the DAMA/LIBRA annual modulation signal as being due to DM scattering off electrons through axial-vector coupling as described in~\cite{Aprile:acpaper}, with a recently refined calculation taking relativistic effects into account~\cite{ermodelnew2}. Fixing the period to 1~year, the best-fit amplitude and phase can be extracted as $(1.67\pm 0.73)$~events/(keV$\cdot$tonne$\cdot$day) and $(136\pm25)$~days respectively. \figref{plphase} shows the confidence level contours from a PL scan. If the DAMA/LIBRA signal were caused by DM scattering off electrons, the expected modulation amplitude in XENON100 would be $(12.2\pm1.2_{\rm{stat}}\pm0.7_{\rm{syst}})$~events/(keV$\cdot$tonne$\cdot$day). Although the modulation phase in the DAMA/LIBRA experiment is consistent with the best fit once the period is fixed to one year instead of using the period preferred by the data, its modulation amplitude is far larger than that observed by XENON100. The XENON100 data disagrees with a signal of the DAMA/LIBRA modulation at $5.7\,\sigma$.

We thank B.M.~Roberts for discussions and providing the calculated wave functions. We gratefully acknowledge support from: the National Science Foundation, Swiss National Science Foundation, Deutsche Forschungsgemeinschaft, Max Planck
Gesellschaft, Foundation for Fundamental Research on Matter, Weizmann Institute of Science, I-CORE, Initial Training Network Invisibles (Marie Curie Actions, PITNGA-2011-289442), Fundacao para a Ciencia e a Tecnologia, Region des Pays de la Loire, Knut and Alice Wallenberg
Foundation, Istituto Nazionale di Fisica Nucleare, and the Laboratori Nazionali del Gran Sasso for hosting and supporting the XENON project.


\end{document}